%% file: main.tex
\documentclass[runningheads]{llncs}
\usepackage{microtype} %

\usepackage{xcolor}
\usepackage{booktabs}

\usepackage{cite} %

\usepackage[T1]{fontenc}
\usepackage{tikz}

\usepackage{float}
\floatstyle{ruled}
\newfloat{listing}{ht}{lop}
\floatname{listing}{Listing}

\usepackage{listings}

\newlength{\PyXLeftMargin}
\newcommand{\lstinput}[1]{%
  \lstinputlisting{listings/#1.py}%
  \raggedleft\footnotesize\texttt{listings/style.tex} missing; use \texttt{lstlistings} for highlighting
}

\newcommand{\pyOp}[1]{\texttt{#1}}
\newcommand{\pyStr}[1]{\texttt{"#1"}}%

\newlength{\FvXLeftMargin}
\usepackage{fancyvrb,xcolor}
\newcommand{\useFancyVrb}{
  \input{listings/style.tex}%
  \renewcommand{\lstinput}[1]{%
    \bgroup{}%
    \fvset{
      numbers=left,
      numbersep=6pt,
      xleftmargin=\FvXLeftMargin,
      commandchars=€,
      vspace=0pt,
    }%
    \renewcommand{\theFancyVerbLine}{%
      \color{gray}\small\arabic{FancyVerbLine}%
    }%
    \small\ttfamily%
    \input{listings/##1.tex}%
    \egroup{}%
  }%
  \renewcommand{\pyOp}[1]{{\ttfamily\PY{o}{##1}}}%
  \renewcommand{\pyStr}[1]{{\ttfamily\PY{l+s+s2}{\PYZdq{}}\PY{l+s+s2}{##1}\PY{l+s+s2}{\PYZdq{}}}}%
}

\IfFileExists{listings/style.tex}{\useFancyVrb{}}{}

\usepackage[hidelinks]{hyperref}

\begin{document}

\title{Generalizing Selfish Mining for DAG Protocols}
\title{Short Paper: Selfish Mining\\Generalized for DAG Protocols}
\title{Short Paper: Generalizing\\Selfish Mining for DAG Protocols}
\title{Short Paper: A Generic MDP for\\Selfish Mining in DAG Protocols}
\title{Generic Selfish Mining MDP for DAG Protocols}
\author{Patrik Keller}
\authorrunning{P. Keller}
\institute{University of Innsbruck, Austria}
\maketitle

\begin{abstract}
  Selfish Mining is strategic rule-breaking to maximize rewards in proof-of-work protocols~\cite{eyal2014MajorityNot} and Markov Decision Processes (MDPs) are the preferred tool for finding optimal strategies in Bitcoin~\cite{sapirshtein2016OptimalSelfish, gervais2016SecurityPerformance} and similar linear chain protocols~\cite{zhang2019LayCommon}.
  Protocols increasingly adopt non-sequential chain structures~\cite{wang2023SoKDAGbased}, for which MDP analysis is more involved~\cite{barzur2020EfficientMDP}.
  To date, researchers have tailored specific attack spaces for each protocol~\cite{sapirshtein2016OptimalSelfish, gervais2016SecurityPerformance, zhang2019LayCommon, barzur2020EfficientMDP, hou2021SquirRLAutomating, keller2023TailstormSecure}.
  Assumptions differ, and validating and comparing results is difficult.
  To overcome this, we propose a generic attack space that supports a wide range of DAG protocols, including Ethereum, Fruitchains, and Parallel Proof-of-Work. %
  Our approach is modular: we specify each protocol as one program, and then derive the Selfish Mining MDPs automatically.

\end{abstract}

\section{Introduction}

In the context of proof-of-work cryptocurrencies, ``Selfish Mining'' entails the deliberate deviation from the protocol to maximize personal rewards, e.\,g., by temporarily withholding blocks.
The first Selfish Mining strategy was proposed by Eyal and Sirer in 2014 against Bitcoin~\cite{eyal2014MajorityNot}.
Using their strategy, attackers controlling more than a third of the system's hash rate can reap more than their fair share of rewards.
Depending on the attacker's communication advantage, the strategy becomes profitable even for weaker attackers.
Soon after, Sapirshtein et al.~\cite{sapirshtein2016OptimalSelfish} and Gervais et al.~\cite{gervais2016SecurityPerformance} employed Markov Decision Processes (MDPs) to find optimal Selfish Mining strategies against Bitcoin.
In 2019, Zhang and Preneel~\cite{zhang2019LayCommon} comprehensively mapped and analyzed various defense mechanisms. %
They assessed different protocols with respect to Selfish Mining (incentive compatibility), chain quality, subversion gain, and censorship susceptibility.
Their findings indicate that all mechanisms considered for improving incentive compatibility fall short on at least one of the other security metrics.
They conclude that no alternative design completely outperforms Bitcoin.

Unfortunately, Zhang and Preneel's~\cite{zhang2019LayCommon} analysis focuses only on Bitcoin-like protocols with linear blockchains.
Many recent protocols, on the other hand, adopt a directed acyclic graph (DAG) structure. %
These protocols do not fit well with the conventional MDP techniques \cite{sapirshtein2016OptimalSelfish, gervais2016SecurityPerformance, zhang2019LayCommon}, since they often take into account the DAG's structure to resolve ties and assign rewards.
However, keeping track of the DAG's structure results in a significant increase in complexity and has kept MDP-based analysis out of reach for DAG protocols.

Fortunately, recent research has provided new avenues for the analysis of complex protocols.
Probabilistic Termination, a technique introduced by Bar-Zur et al. in 2020~\cite{barzur2020EfficientMDP}, enables the \emph{exact} solution of traditional selfish mining MDPs with a tenfold efficiency improvement.
When this approach meets its limits, one can turn to \emph{approximating} reinforcement learning techniques, as suggested by Hou et al. in 2021~\cite{hou2021SquirRLAutomating} and later applied by Keller et al.~\cite{keller2023TailstormSecure} and Bar-Zur et al.~\cite{barzur2022WeRLmanTackle}.

We observe that all mentioned attack searches~\cite{sapirshtein2016OptimalSelfish, gervais2016SecurityPerformance, zhang2019LayCommon, barzur2020EfficientMDP, hou2021SquirRLAutomating, keller2023TailstormSecure} tailor individual MDPs for each protocol analyzed.
Individual MDPs embed a wide range of assumptions about what the attacker can do and how the other miners react.
Subtle differences can have significant effects which makes results difficult to compare. %
E.\,g., one source of confusion is that researchers tend to consider different versions (or subsets) of Ethereum~\cite{zhang2019LayCommon, barzur2020EfficientMDP, hou2021SquirRLAutomating}.

We present ongoing work aimed at addressing this issue.
Our approach is modular: we explicitly specify each protocol once before applying multiple security analyses automatically.
This facilitates the fair comparison of protocols using a uniform set of assumptions and enables scaling the number of protocols.
Using the terminology of Wang et al.~\cite{wang2023SoKDAGbased}, we support the class of proof-of-work DAG protocols that provide a total ordering of blocks.
In this short paper, we focus on Selfish Mining and validate against established results for Bitcoin~\cite{sapirshtein2016OptimalSelfish}.
Future versions will add more security metrics~\cite{zhang2019LayCommon} and protocols. %

We proceed as follows:
Section~\ref{sec:traditional} recapitulates the traditional Selfish Mining model for Bitcoin;
Section~\ref{sec:proposed} presents our generic model;
Section~\ref{sec:validation} validates our model against the traditional model; and
Section~\ref{sec:outlook} gives a short outlook.

\section{Traditional Selfish Mining against Bitcoin}\label{sec:traditional}

The traditional Selfish Mining attack space~\cite{sapirshtein2016OptimalSelfish} models two nodes:
the attacker and the defender.
The attacker can fork the public chain, mine blocks in private, and selectively release withheld blocks later.
The defender follows the protocol as specified.
This is pessimistic in that it assumes multiple attackers will collude.
The model is limited to one fork and there are at most two chains:
the attacker grows the private chain and the defender grows the public chain.

Traditionally, the attacker may choose from the four actions: \emph{Adopt}, \emph{Match}, \emph{Override}, and \emph{Wait}.
The \emph{Adopt} action implies discarding the private chain and starting a new fork from the most recent public block.
With \emph{Match} the attacker releases just enough blocks to induce a block race between the public and the released subset of the private chain.
With probability $\gamma$ the block race is resolved in favor of the attacker, that is, the defender discards the public chain and continues mining on the released subset of the private chain.
The \emph{Override} action is similar to \emph{Match}, but the attacker releases one additional block.
This forces the defender to discard the recent blocks of the public chain.
Lastly, the \emph{Wait} action lets the attacker continue mining on the private chain without doing anything.

Note how $\gamma$ models the communication advantage of the attacker.
Honest Bitcoin nodes resolve block races in favor of the block first received.
High $\gamma$ implies that the attacker can, in reaction to an honest node mining and sending a block, deliver his own block to the other defenders before they learn about the competing block.
A separate parameter, $\alpha$, models the attacker's mining power, or more precisely, the relative frequency of blocks mined on the private chain as opposed to the public chain.

An important benefit of this Selfish Mining model is its simple state space when implemented as an MDP:
two non-negative integers describing the length of the two chains and a third variable tracking the feasibility of \emph{Match} are sufficient.
But this model is naturally restricted to Bitcoin-like protocols that internally only count blocks.
DAG-based protocol features often require tracking many more properties and, hence, more complex state spaces.
For example, Ethereum discounts \emph{uncle block} rewards based on the depth at which the uncles forked off from the main chain.
Others have expended substantial effort tailoring an MDP for this protocol feature~\cite{barzur2020EfficientMDP}.
The approach we take is to identify the relevant parts of the DAG and then track its entire structure in the MDP's state.
This allows us to cover many protocols in a single model.

\section{Generic Attack Space for DAG Protocols}

In deployed proof-of-work protocols, a block is a message with a rare cryptographic hash value.
Tweaking a message to satisfy this property, and thereby create a valid block, requires seconds to minutes of repeated hashing by the entire network.
Invalid blocks can easily be rejected by checking their hash value; hence, proof-of-work effectively rate-limits participation in the system.

Blocks can refer to other blocks by including their hashes.
If block \texttt{A} refers to block \texttt{B}, then we say that \texttt{B} is a parent of \texttt{A} and that \texttt{A} is a child of \texttt{B}.
The parent relationship spans a directed acyclic graph (DAG) of blocks (BlockDAG).
The protocol specification defines how honest nodes grow the BlockDAG, how the difficulty adjustment algorithm (DAA) should adjust the hash threshold, and how much reward is assigned to each miner.

\subsubsection{Protocol Specification}

The primary objective of proof-of-work consensus protocols is to order cryptocurrency transactions and ensure that this ordering is not changed after the fact.
Typically, each block contains a pre-ordered list of transactions, while the protocol defines a unique predecessor for each block.
In combination, each block defines an unambiguous sequential history of transactions.
Participating nodes typically maintain a pointer to their preferred block.
Thereby, they also define their preferred version of the transaction history.
The DAA inspects the history of the preferred block and, in response, sets the hash threshold for future blocks.
Reward allocations are cryptocurrency transactions, hence they depend on the preferred block as well.
Based on these observations, we propose to specify protocols as a set of five pure functions:
\begin{description}
  \item{\texttt{mining(B)}} prescribes how miners grow the BlockDAG.
    The function takes as argument the currently preferred block \texttt{B} and returns the set of blocks that should be referenced in the next mined block.

  \item{\texttt{update(A, B)}} defines how miners update their preferred block after they have mined or received a new block.
    The function takes as arguments the currently preferred block \texttt{A} and a new block \texttt{B} and returns the block to be preferred in the future.
    The function may return more than one block to model uniform tie breaking as proposed by Eyal and Sirer~\cite{eyal2014MajorityNot}.

  \item{\texttt{previous(B)}} returns one \emph{ancestor} of block \texttt{B} or nothing.
    Calling the function recursively yields the sequential history of block \texttt{B}.
    Note that the returned block does not have to be a direct \emph{parent} of \texttt{B}.

  \item{\texttt{progress(B)}} returns a real number defining the size of the sequential history of \texttt{B}.
    The DAA will adjust the hash value threshold for new blocks such that the history grows at a pre-defined rate.
    It creates a feedback loop between the system's overall hash rate, progress, and time.

  \item{\texttt{coinbase(B)}} defines the cryptocurrency rewards minted by block \texttt{B}.
    The function returns a list of tuples $(r, v)$ where $r$ denotes the recipient and $v$ the value of the reward.
    We evaluate this function for all blocks in the sequential history and accumulate rewards accordingly.
\end{description}
The protocol specification may access the BlockDAG as follows.
\begin{description}
  \item{\texttt{parents(B)}} returns the parents of block \texttt{B}.
  \item{\texttt{children(B)}} returns the children of block \texttt{B}.
  \item{\texttt{miner(B)}} returns the miner of block \texttt{B}.
  \item{\texttt{height(B)}} returns the length of the shortest path from \texttt{B} to a root block.
\end{description}
Listing~\ref{alg:bitcoin} provides an example specification for the Bitcoin protocol in Python.

\begin{listing}[t]
  \caption{Specification of the Bitcoin Protocol}
  \label{alg:bitcoin}
  \lstinput{bitcoin}
\end{listing}

\subsubsection{Attack Space}\label{sec:implicit}\label{sec:proposed}

\input{fig/state.tex}

We now adapt the traditional Selfish Mining attack space from Section~\ref{sec:traditional} to the generic DAG protocol specified as described above.
We copy the model parameters $\alpha$ (relative mining power) and $\gamma$ (attacker communication advantage).
We also model only two system participants, attacker and defender.
The defender follows the protocol as specified, the attacker may choose from predefined actions.
Unlike before, where the model's state was composed of two counters and a third variable, we now track the structure of the entire BlockDAG.

We observe that the actions of the traditional model for Bitcoin control the withholding of private blocks and the ignoring of public blocks.
A-priori, newly mined attacker blocks are withheld and new defender blocks are ignored, resulting in a growing fork (Wait action).
The Match and Override actions cause the release of previously withheld blocks.
The Adopt action captures the attacker's capitulation to previously ignored public blocks.

Based on this observation, we define generic actions for \emph{releasing} and \emph{considering} individual blocks of the DAG.
A third action is for \emph{continuing} mining withheld private blocks and ignoring public blocks, respectively.
It's important to note that the attacker behaves honestly apart from withholding and ignoring blocks.
In particular, the attacker will grow the DAG according to the protocol's \texttt{update} and \texttt{mining} rules but will consider only a subset of the BlockDAG.

\subsubsection{State}

The MDP's state consists of a BlockDAG where each block has three properties that separately track what the defender knows, what the attacker ignores, and what the attacker withholds.
Figure~\ref{fig:state} shows an example.

Each block can be either \texttt{Unknown}, \texttt{Known}, or \texttt{PreferredD} by the defender.
New blocks are initially \texttt{Unknown}.
Exactly one block is \texttt{PreferredD} at any time.
Blocks marked \texttt{Known} or \texttt{PreferredD} now will not be marked \texttt{Unknown} in the future.
If a block is marked \texttt{Known} or \texttt{PreferredD} then no parent may be marked \texttt{Unknown}.
In other words, the defender learns about blocks in topological order.

Each block can be either \texttt{Ignored}, \texttt{Considered}, or \texttt{PreferredA} by the attacker.
Blocks are \texttt{Ignored} initially.
Exactly one block is \texttt{PreferredA} at any time.
Blocks marked \texttt{Considered} or \texttt{PreferredA} now will not be marked \texttt{Ignored} in the future.
If a block is marked \texttt{Considered} or \texttt{PreferredA} then no parent may be marked \texttt{Ignored}, i.\,e., the attacker considers blocks in topological order.

Each block can be either  be \texttt{Foreign}, \texttt{Withheld}, or \texttt{Released}.
Blocks mined by the defender are \texttt{Foreign} and will stay so forever.
Blocks mined by the attacker are initially \texttt{Withheld} and may eventually be \texttt{Released}.
Blocks marked \texttt{Released} now cannot be \texttt{Withheld} in the future.
The attacker releases blocks in topological order.

We define two start states.
In both cases, the DAG holds a single block, which we call \emph{genesis} or \emph{root}. Both the attacker and defender prefer the genesis.
With probability $\alpha$, we mark the genesis \texttt{Released} implying the block was mined by the attacker.
Otherwise, with probability $1 - \alpha$, we mark the genesis as \texttt{Foreign}.

\subsubsection{Actions}

We define the three actions \texttt{Release(i)}, \texttt{Consider(i)}, \texttt{Continue}.
They induce probabilistic state transitions as follows.

\texttt{Release(i)} shares an individual block with the defender.
Among all candidate blocks, that is \texttt{Withheld} blocks where no parent is \texttt{Withheld}, we mark the \texttt{i}-th block \texttt{Released}.
This state transition is deterministic.
The action is infeasible if there are fewer than $i$ candidate blocks.

With \texttt{Consider(i)} the attacker stops ignoring an individual block.
Among all candidate blocks, that is \texttt{Ignored} blocks where no parent is \texttt{Ignored}, we mark the \texttt{i}-th block \texttt{Considered}.
We then call the protocol's \texttt{update} function with the attacker's old preferred block and the block to be considered while hiding the \texttt{Ignored} subset of the BlockDAG.
The new preferred block is recorded by setting \texttt{PreferredA} accordingly.
This state transition is deterministic.
The action is infeasible if there are fewer than $i$ candidate blocks.

The \texttt{Continue} action proceeds in time until the next block is mined.
The action is always feasible.
It has four outcomes, depending on whether the attacker communicates quickly (probability $\gamma$) and whether she mines the next block (probability $\alpha$).
If the attacker communicates quickly, we find all blocks that are both \texttt{Released} and \texttt{Unknown} and deliver them to the defender in topological order.
We deliver individual blocks by marking the block \texttt{Known}, evaluating the protocol's \texttt{update} rule while hiding \texttt{Unknown} blocks, and marking the returned block \texttt{PreferredD}.
Subsequently, we find the blocks that are both \texttt{Foreign} and \texttt{Unknown} and deliver them to the defender as just described.
Otherwise, if the attacker communicates slowly, we proceed in the opposite order and deliver the \texttt{Foreign} blocks before the \texttt{Released} ones.

If the attacker mines the next block, we evaluate the protocol's \texttt{mining} function for the block marked \texttt{PreferredA} while hiding \texttt{Ignored} blocks.
We extend the DAG accordingly, setting the new block to \texttt{Unknown}, \texttt{Withheld} and \texttt{Ignored}.
Otherwise, if the defender mines the next block, we evaluate the \texttt{mining} function for the block marked \texttt{PreferredD} while hiding \texttt{Unknown} blocks.
We append a new block accordingly and set it to \texttt{Unknown}, \texttt{Foreign} and \texttt{Ignored}.

Reconsidering the state in Figure~\ref{fig:state}, we now see that it can occur in Bitcoin:
the defender has just mined a block of height two and the attacker attempts to \emph{Match} by releasing a competing block of the same height.

\subsubsection{State Space Compression}\label{sec:explicit}

The generality of the DAG-based state space comes at the price of increased complexity.
We employ multiple tricks to avoid redundancy and keep the size of the state space manageable.

First, we observe that the state space, as described above, is infinite.
The \texttt{Continue} action is always feasible and it always appends a block.
We overcome this by setting an artificial limit, as is common in related work~\cite{sapirshtein2016OptimalSelfish, gervais2016SecurityPerformance, barzur2020EfficientMDP, hou2021SquirRLAutomating}:
mining new blocks is only possible until the BlockDAG reaches a certain size (or height).

Second, we explore only those states that are reachable from the two well-defined start states using feasible actions.
This takes into account the given protocol specification.
For example, in the Bitcoin protocol, each state will describe a tree of blocks, while more general DAGs are excluded.

Third, we observe that even when attacker and defender prefer different blocks, they still agree on a common prefix of the sequential history. %
The common history initially consists of the genesis.
It grows as the attacker considers or releases blocks.
We assert that the common history is irrelevant to future actions.
This allows us to discard all ancestors of the latest common ancestor.
After truncation, the latest common ancestor becomes the new genesis block.

Fourth, we avoid isomorphic states by storing the BlockDAG in a canonical form.
For this, we translate the tuple of the defender's view, ignoring status, and withholding status (see Fig.~\ref{fig:state}) into $3^3 = 27$ vertex colors.
We then find a canonical and color-preserving vertex labeling using appropriate third-party software, Nauty~\cite{mckay2014PracticalGraph}.
This merges states that are equal apart from the chronological order in which blocks were mined.
E.\,g., the 6 blocks of the example state depicted in Figure~\ref{fig:state} could be ordered in 10 different ways.

\subsubsection{Rewards and Progress}\label{sec:reward}

Recall that we truncate the common history after each probabilistic state transition.
Just before removing blocks from the DAG, we iterate through the sequential history of the latest common ancestor using the protocol's \texttt{previous} function.
For each block returned, we evaluate the \texttt{coinbase} function.
Depending on the protocol, it might return zero, one, or multiple reward assignments, each benefiting either the attacker or the defender.
We accumulate the attacker's and defender's rewards separately and record them together with the probabilistic state transition.
We proceed similarly for the protocol's \texttt{progress}.

\section{Validation}\label{sec:validation}

We implement two versions of the traditional model \cite{sapirshtein2016OptimalSelfish, barzur2020EfficientMDP} (Section~\ref{sec:traditional}) and the proposed model (Section~\ref{sec:proposed}).
We instantiate our model for Bitcoin (Listing~\ref{alg:bitcoin}).
We then apply the same Selfish Mining analysis to all three models.
This isolates the effect of the model from the peculiarities of policy optimization and evaluation.

The main principle behind Selfish Mining is maximizing long-term revenue.
We assume the attacker's mining expenses are linear in time and that the attacker's earnings correspond one-by-one to the cryptocurrency rewards assigned by the protocol.
Recall that the DAA adjusts the block mining difficulty such that the sequential history grows at a given rate.
This introduces a feedback loop between the chosen policy, time, and the protocol's \texttt{progress} function.
As a result, long-term revenue becomes non-linear and depends on the entire history of the process~\cite{sapirshtein2016OptimalSelfish}.
Standard MDP solving techniques cannot be applied.

An effective workaround is Probabilistic Termination (PT)~\cite{barzur2020EfficientMDP}, which modifies the MDP such that any policy terminates after a certain amount of progress has been made.
Assuming the DAA has stabilized with respect to the policy, constant expected progress implies constant expected time.
After applying PT, revenue-maximizing policies can be identified using standard algorithms. %

\begin{figure}[t]
  \includegraphics[width=\linewidth]{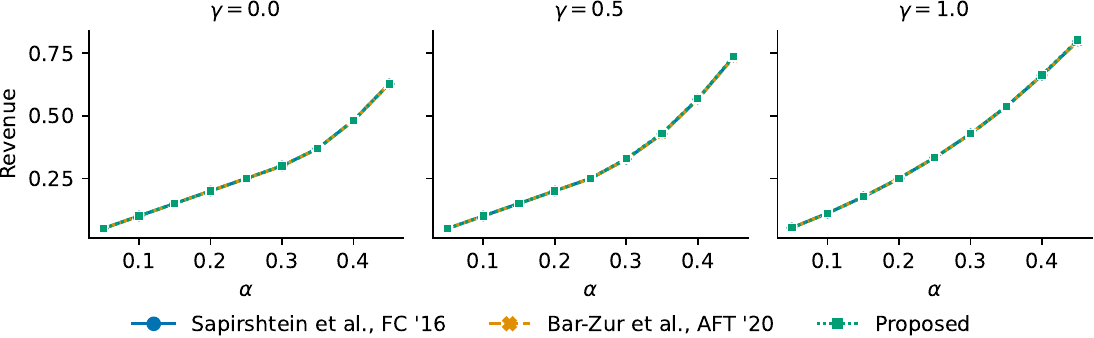}
  \caption{
    Selfish Mining revenue against Bitcoin for varying communication advantage~$\gamma$ (column) and mining power~$\alpha$ (x-axis).
    We compare our model against the traditional approach presented in Section~\ref{sec:traditional}.
    Our results show perfect alignment with Sapirshtein et al.~\cite{sapirshtein2016OptimalSelfish} (blue line) and Bar-Zur et al.~\cite{barzur2020EfficientMDP} (orange line).
  }
  \label{fig:validation}
\end{figure}

\bgroup \input{fig/tab-validation-meta.tex} %

Figure~\ref{fig:validation} plots the Selfish Mining revenue (y-axis) for varying mining power~$\alpha$ (x-axis), attacker communication advantage $\gamma$~(facet), and the three competing models (style and color).
All models truncate the attack space at height $\maximumHeight$.
The policies are obtained from applying PT with expected progress $\horizon$ and value iteration stopping at revenue delta $\varepsilon = \eps$.
We evaluate the policies in the source model (before PT).
The y-axis reports expected reward per progress.

We observe that our generic model (instantiated for Bitcoin) yields results identical to the traditional models (specific to Bitcoin), as described by Sapirshtein et al.~\cite{sapirshtein2016OptimalSelfish} and Bar-Zur et al.~\cite{barzur2020EfficientMDP}.

\egroup %

\section{Outlook}\label{sec:outlook}

Besides Bitcoin, we have specified simplified versions of parallel proof-of-work~\cite{keller2022ParallelProofofwork} and Tailstorm~\cite{keller2023TailstormSecure}, and two variants of Ethereum (Whitepaper and Byzantium).
In addition to long-term Selfish Mining revenue, we plan to model short-term revenue, censoring, and history rewriting.
Reinforcement learning can be applied to overcome the rapidly expanding state space while growing the size limit of the BlockDAG~\cite{hou2021SquirRLAutomating, keller2023TailstormSecure}.
Equally interesting would be the exact modeling of difficulty adjustment algorithms~\cite{negy2020SelfishMining} and the incorporation of transaction fees~\cite{barzur2022WeRLmanTackle}.

\section*{Acknowledgements}
I express my gratitude to my colleagues, George Bissias and Michael Fröwis, for reviewing this work and offering valuable suggestions.

\bibliographystyle{splncs04}
\bibliography{do-not-edit-this-file-manually.bib, do-edit-this-one.bib}

\end{document}

%% file: fig/state.tex
\begin{figure}
  \centering
  \newcommand{\block}[5]{
    \begin{scope}[shift={#1}]
      \draw (-1.7em,0) node[draw,rectangle,minimum width=1.7em] {\vphantom{I}#3};
      \draw (0,0)      node[draw,rectangle,minimum width=1.7em] {\vphantom{I}#4};
      \draw (1.7em,0)  node[draw,rectangle,minimum width=1.7em] {\vphantom{I}#5};
      \node[rectangle, minimum width=5.1em, outer sep=2pt] at (0,0) (#2) {\vphantom{A}};
    \end{scope}
  }
  \begin{tikzpicture}[y=0.6cm, x=10em, >=stealth]
    \block{(0, 0)}{g }{K}{C}{F}
    \block{(1, 1)}{a1}{U}{C}{R}
    \block{(2, 1)}{a2}{U}{C}{R}
    \block{(3, 1)}{a3}{U}{P}{W}
    \block{(1,-1)}{d1}{P}{I}{F}
    \block{(2,-1)}{d2}{U}{I}{F}
    \draw[<-] (g ) -- (a1);
    \draw[<-] (g ) -- (d1);
    \draw[<-] (d1) -- (d2);
    \draw[<-] (a1) -- (a2);
    \draw[<-] (a2) -- (a3);
  \end{tikzpicture}

  \begin{tikzpicture}[y=4ex, x=12em]
    \node[right] at ([xshift=-2.9em] 0, 3.75) {\textbf{Legend}};
    \node[right] at ([xshift=-2.9em] 0, 3) {Defender's View};
    \block{(0, 2)}{dU}{U}{}{}
    \block{(0, 1)}{dK}{K}{}{}
    \block{(0, 0)}{dP}{P}{}{}
    \node[right] at ([xshift=2.5em]dU) {\ttfamily Unknown};
    \node[right] at ([xshift=2.5em]dK) {\ttfamily Known};
    \node[right] at ([xshift=2.5em]dP) {\ttfamily PreferredD};

    \node[right] at ([xshift=-2.9em] 1, 3) {Ignoring Status};
    \block{(1, 2)}{aI}{}{I}{}
    \block{(1, 1)}{aC}{}{C}{}
    \block{(1, 0)}{aP}{}{P}{}
    \node[right] at ([xshift=2.5em]aI) {\ttfamily Ignored};
    \node[right] at ([xshift=2.5em]aC) {\ttfamily Considered};
    \node[right] at ([xshift=2.5em]aP) {\ttfamily PreferredA};

    \node[right] at ([xshift=-2.9em] 2, 3) {Withholding Status};
    \block{(2, 2)}{wF}{}{}{F}
    \block{(2, 1)}{wW}{}{}{W}
    \block{(2, 0)}{wR}{}{}{R}
    \node[right] at ([xshift=2.5em]wF) {\ttfamily Foreign};
    \node[right] at ([xshift=2.5em]wW) {\ttfamily Withheld};
    \node[right] at ([xshift=2.5em]wR) {\ttfamily Released};
  \end{tikzpicture}
  \caption{
    Example state with $6$ blocks.
    The arrows indicate the \texttt{parent} relationship.
    The leftmost block is the genesis block.
    Three variables per block track what the defender knows and what the attacker ignores and withholds.
    This state may occur in Bitcoin; in general, blocks may have more than one parent, e.\,g., uncle blocks in Ethereum.
  }
  \label{fig:state}
\end{figure}

%% file: fig/tab-validation-meta.tex
%
\def\maximumHeight{13}
\def\ourStates{159819}
\def\ourActions{4}
\def\ourTransitions{846080}
\def\fcStates{417}
\def\fcActions{4}
\def\fcTransitions{2362}
\def\horizon{100}
\def\eps{0.01}